\documentclass[twocolumn]{jpsj2}

\usepackage{bm}

\title{Effect of Quadratic Zeeman Energy on the Vortex of Spinor Bose-Einstein Condensates}
\author{Tomoya Isoshima and Sungkit Yip}
\inst{Institute of Physics, Academia Sinica, Nankang, Taipei 11529, Taiwan}

\date{\today}

\abst{
The spinor Bose-Einstein condensate of atomic gases
has been experimentally realized by a number of groups.
Further, theoretical proposals of the possible vortex states
have been sugessted.
This paper studies the effects of the quadratic Zeeman energy
on the vortex states.
This energy was ignored in previous theoretical studies,
although it exists in experimental systems.
We present phase diagrams of various vortex states
taking into account the quadratic Zeeman energy.
The vortex states are
calculated by the Gross-Pitaevskii equations.
Several new kinds of vortex states are found.
It is also found that
the quadratic Zeeman energy affects the direction of total magnetization
and
causes a significant change in the phase diagrams.
}

\kword{Vortex, Spinor, Bose-Einstein condensate}

\begin{document}
\maketitle

\section{Introduction}

Quantized vortices in the
Bose-Einstein condensate (BEC) of ultracold atoms
have been a topic of active research.
Since the first realization of the vortex~\cite{matthews1999},
a variety of studies have been carried out
such as those on the formation of the vortex lattice,
bending and oscillation of vortex lines~\cite{bretin2003},
vortex rings, and so on.
For a review of the early studies, see ref.~\citen{fetter2001}.
The interest in vortices in ultracold atoms is
also increasing with regard to more complicated systems, such as
the BEC interacting with an optical lattice,
BEC with dipolar interactions~\cite{griesmaier2005,cooper2005,zhang2005}, and
molecules and Cooper pairs of Fermions~\cite{zwierlein2005}.

Despite these extensive studies on vortices,
studies on simpler systems such as the spinor BEC of atoms
in a uniform magnetic field~\cite{yip1999,isoshima2002,mizushima2002}
remains unachieved.
The existence of various vortices has been predicted in
the seminal papers by Ho \cite{ho1998} 
and Ohmi and Machida \cite{ohmi1998}
and was subsequently continued by many works~\cite{yip1999,isoshima2002,mizushima2002,martikainen2002,kobayashi2004,mueller2004}.

In the spinor BEC, a vortex state 
is inseparably related to the spin texture.
By utilizing this relation, 
it is possible to imprint vorticity without rotating the cloud
as proved in the experiments~\cite{leanhardt2003}.
By applying a quadrupole (nonuniform) magnetic field,
the spin texture of the BEC is forced,
resulting in the nucleation of the vortex.

Conversely,
for a rotated spinor BEC in a uniform magnetic field or
without a magnetic field,
many theoretical predictions on the possible spin textures 
have been made~\cite{yip1999,isoshima2002,kobayashi2004,martikainen2002,mueller2004}.
In this case, the rotation of the trapping potential causes
vorticity that produces a spin texture.
In a spinor BEC in a uniform field,
the boundary condition and magnetic field gradient
do not explicitly control the vorticity.
Instead, the internal energy of the vortices determines the
vortex configuration.
However, the experimental observation of vortices still remains to be performed.

In an optically confined BEC,
a dominant effect ---the linear Zeeman effect--- due to
the intensity of the magnetic field
is cancelled by the conservation of
the total magnetization (see ref.~\citen{chang2004}).
Nevertheless, the magnetic field intensity affects the internal spin
through the quadratic Zeeman effect,
which was ignored in previous studies~\cite{yip1999,isoshima2002}
on the vortices in the spinor BEC.
The quadratic Zeeman energy is proportional to
the square of the magnetic field intensity and the local spin.

In this paper, 
we take into account the quadratic Zeeman energy
and present new phase diagrams of vortices
calculated by the extended Gross-Pitaevskii equation.
It is important to consider the quadratic Zeeman energy
because its existence in reality is inevitable.
In addition, the quadratic Zeeman energy leads to vortex reconfiguration 
within fixed values of magnetization $M$ and angular momentum $L$.

Atomic interactions are classified into two types:
ferromagnetic and antiferromagnetic (polar).~\cite{ho1998,ohmi1998}
In this study, we treat the condensate of sodium atoms ($F=1$),
which is antiferromagnetic. 
In the absence of a magnetic field for antiferromagnetic systems, 
vortices (and their lattices)
in which the up and down components are populated
are stable.~\cite{martikainen2002,mueller2004}
They have been labelled as the
vortex of region II~\cite{yip1999} or
the $(1,\times, 0)$ type~\cite{isoshima2002,mizushima2002}.
In a magnetic field, or in the presence of linear Zeeman energy,
various other vortex states are found to be possible.
The present work aims to study
the effects of the quadratic Zeeman energy on such vortex states
and to extend the phase diagrams.

\section{Model}

We simulate the vortex states of the BEC of atomic gases
by numerically solving the extended Gross-Pitaevskii equation.
The system is rotated with an angular velocity $\Omega$,
and a uniform magnetic field is applied.
The calculation 
is similar to that in ref.~\citen{yip1999},
but we take into account the quadratic Zeeman energy.
For simplicity, we treat a harmonically trapped
two-dimensional system.


We assume the direction of the applied magnetic field 
and the direction of quantization axis to be the $z$-axis.
The vector of the rotation $\Omega$ is also in the $+z$ direction.

The order parameter of the condensate consists of macroscopic
wavefunctions $\Psi_j$,
where $j = 0,\pm1$ are hyperfine projections.
The order parameters $\Psi_j$ are given by the minimization of a free energy
\begin{equation}
G = E_1 +  E_\mathrm{int} + E_\mathrm{Zee} - \Omega L - \mu N,
\end{equation}
which consists of the single particle energy $E_1$, 
the interaction energy $E_\mathrm{int}$, and
Zeeman energy $E_\mathrm{Zee}$; they are
\begin{eqnarray}
    E_{1} &=& \sum_j \int d{\bm r}
    \Psi_j^\ast \left( 
        -\frac{\hbar^2}{2M_a}\nabla^2 
    \right) \Psi_j + V(\mathbf{r}) n_j
\\
    E_\mathrm{int} &=& \int d{\bm r}\left(
        \frac{c_0 + c_2}{2} n^2 -
        \frac{c_2}{2} |2 \Psi_1\Psi_{-1} - \Psi_0^ 2|
    \right),
\label{eq:e:int}
\\
    E_\mathrm{Zee} &=& \sum_j \int d{\bm r} \left( -p j n_j + q j^2 n_j \right),
\label{eq:e:zee}
\end{eqnarray}
respectively,
where the densities $n_j = |\Psi_j|^2$ and $n = \sum_j n_j$,
particle number $N = \sum_j \int n_j({\bm r}) d{\bm r}$,
angular momentum $L = - \mathrm{i}\hbar \sum_j \int \Psi_j^\ast \nabla \Psi_j \times {\bm r} d{\bm r}$,
spin index $j \in {1, 0, -1}$, and
trapping potential $V(\mathbf{r}) = \frac{m(2\pi \nu_r)^2}{2}(x^2 + y^2)$.
The term proportional to $p$ ($q$) in eq.~(\ref{eq:e:zee}) 
represents the linear (quadratic) Zeeman energy. 

The minimization of $G$ gives 
the established Gross-Pitaevskii (GP) equation for
the spinor BEC~\cite{ho1998,ohmi1998}.
We solve the GP equation numerically for various 
values of the quadratic Zeeman factor $q$ to investigate its effect
on the vortex configurations.
The parameters $p$ and $\mu$ are varied
to obtain results with various magnetization $M_z$ at fixed $N=N_\mathrm{2D}$.

The particle number of the $j$-th component is 
$N_j \equiv \int n_j({\bm r}) d{\bm r}$.
The normalized angular momentums of the $j$-th component 
$L_j \equiv - \mathrm{i} \hbar \sum_j \int \Psi_j^\ast \nabla \Psi_j \times {\bm r} d{\bm r} / N_j$
where the spin index $j \in {1, 0, -1}$ are extensively
used to identify vortices.
The local and total magnetizations are
$m_\alpha(\mathbf{r}) = \Psi^\dagger(\mathbf{r}) F_\alpha \Psi(\mathbf{r})$
and
$M_\alpha = \int m_\alpha(\mathbf{r}) d\mathbf{r}$,
respectively, 
where $\alpha = x,y,z$.


A system parameter $\varepsilon$ is defined as follows.
From the mass $M_a$ of atom,
an effective scattering length $a = (a_0 + 2a_2)/3$, 
the interaction parameter $c_0 = 4 \pi \hbar^2 a / M_a$ and
harmonic trap frequency $\omega_0$, 
we can calculate a harmonic oscillator length
 $\lambda_0 = \sqrt{ \hbar / M_a \omega_0 }$.
We use a ratio of oscillator length to condensate radius $R_\mathrm{TF}$,
$\varepsilon = (\lambda_0 / R_\mathrm{TF})^2$ as a system parameter.
Here the radius $R_\mathrm{TF}$ is that of condensate
within Thomas-Fermi (TF) approximation in the absence of spin interaction.
For the system parameter,
values $\varepsilon = 0.1 \textrm{ and } 0.01$ are used.
After a $\varepsilon$ is given, 
we have a particle number
$N_\mathrm{2D} = 1 / (16 \epsilon^2 a)$
within the TF approximation.
We fix the particle number $N$ to $N_\mathrm{2D}$.
The approximate peak density becomes $n_\mathrm{peak} = h \nu_r / (2 c_0 \varepsilon)$
where  $h$ is Plank's constant.  
For sodium atom when $\nu_r = 100 \, \textrm{Hz}$,
$n = 0.329 \, (3.29) \times 10^{14} \textrm{cm}^{-3}$
for $\varepsilon = 0.1 \, (0.01)$.

Value $\varepsilon = 0.1$ is used for consistency with our previous study.~\cite{yip1999}
Another value $\varepsilon = 0.01$ is employed
to simulate thin disk-shaped condensate.
If we assume tight harmonic confinement and
Gaussian shape of condensate wavefunction along $z$-axis,
the particle number in a system is
$N_\mathrm{3D} = \sqrt{ \frac{\hbar \pi}{M_a \omega_z}}N_\mathrm{2D}$
where $\omega_z$ is the confinement parameter along $z$-axis.
For sodium atoms when $\omega_z = 1000 \times 2 \pi$ and $\epsilon = 0.01$,
$N_\mathrm{3D}$ is $2.65 \times 10^5$.
This is within typical experimental particle number.

In an experiment~\cite{stenger1998} without a vortex,
$q$ ranges from $0 \textrm{ to } 70 \, \textrm{Hz}$ for magnetic fields $B_0$
up to $500 \, \textrm{mG}$.
We choose $q/(h \nu_r)$ from 0 to 0.1. The maximum 0.1 corresponds to
$190 \, \mathrm{mG}$ when $\nu_r = 100 \mathrm{Hz}$ and
$q \propto B_0^2$.


The parameters that we explicitly give its values
are the system parameter $\varepsilon$,
the angular velocity $\Omega$ and the quadratic Zeeman factor $q$.
%
%
The total magnetization $M_z$ and the particle number $N$
are controlled via the linear Zeeman factor $p$ and chemical potential $\mu$.
We employ the ratio of the interaction parameters $c_2/c_0 = 0.02$,
which is obtained from an experiment~\cite{stenger1998}
that directly treats the quadratic Zeeman energy and interaction parameter.
Similar to that experiment, we assume the condensate of sodium atoms with $F=1$.

\begin{figure}
\begin{center}
\includegraphics[width=7.5cm]{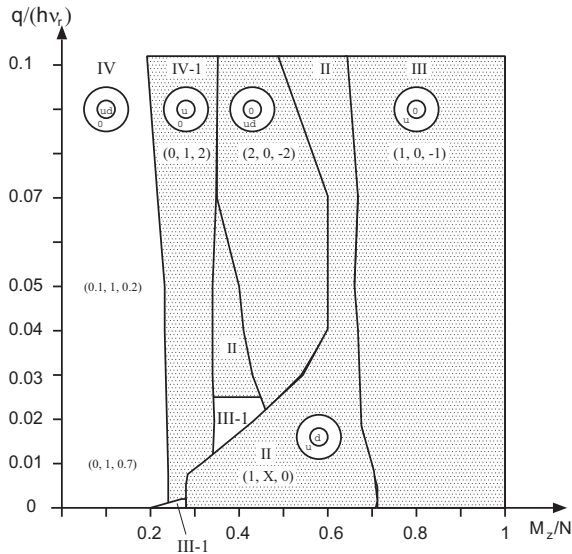}\\
(a)\\
\quad\\
\includegraphics[width=7.5cm]{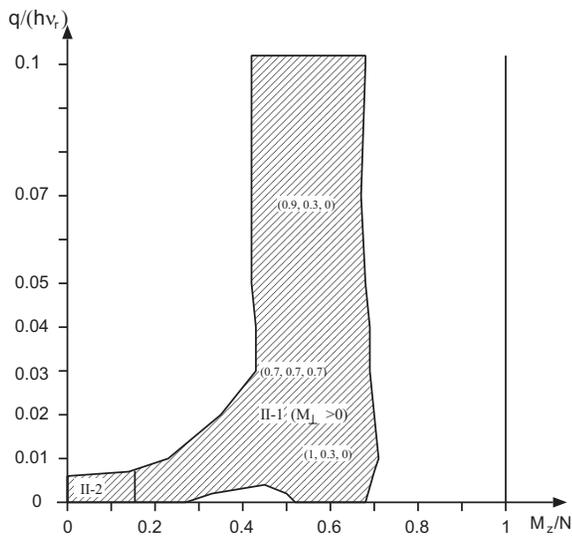}\\
(b)\\
\end{center}
\caption{\label{fig:eps0.1}
Phase diagrams of magnetization $M_z/N$ vs
the quadratic Zeeman term parameter $q/(h\nu_r)$
at $\epsilon=0.1$ and $\Omega = 0.45$.
(a)
This diagram shows the areas for the vortices with $M_\perp = 0$.
A set of three numbers in parentheses \textit{e.g.}, $(1,0,-1)$
represents the angular moments $L_1$, $L_0$ and $L_{-1}$
of the condensate, respectively.
In a shaded region, when the numbers in the set are integers,
all the vortex states in that region has that set of angular moments.
Otherwise, a set of numbers shows only one typical value in the region.
Letters u, 0, or d in double circles 
indicates that the up, zero, or down species is dominant 
inside (inner circle) or outside (outer circle) the centered vortex core.  
(b)
Regions II-1 and II-2 in which vortex states with $M_\perp \ne 0$
have a lower energy are shown.
The perpendicular magnetization $M_\perp / N$ in II-1 lies
between $0.001 \sim 0.2$ at $q/\nu_r > 0$.
At $q=0$ and $M_z/N > 0.5$ within II-1, $M_\perp / N$ is as low as $10^{-8}$.
}
\end{figure}

There are a number of vortex states at each $M_z$.
Among them, we assume that a state with the lowest energy $G$ is realized.
These lowest energy states are presented in the phase diagrams 
---Figs.~\ref{fig:eps0.1}(a), \ref{fig:eps0.1}(b) and \ref{fig:eps0.01}.
However, the direction of the total magnetization requires further explanation
before the details of the phase diagrams are examined.

\begin{figure}
\begin{center}
\includegraphics[width=7.4cm]{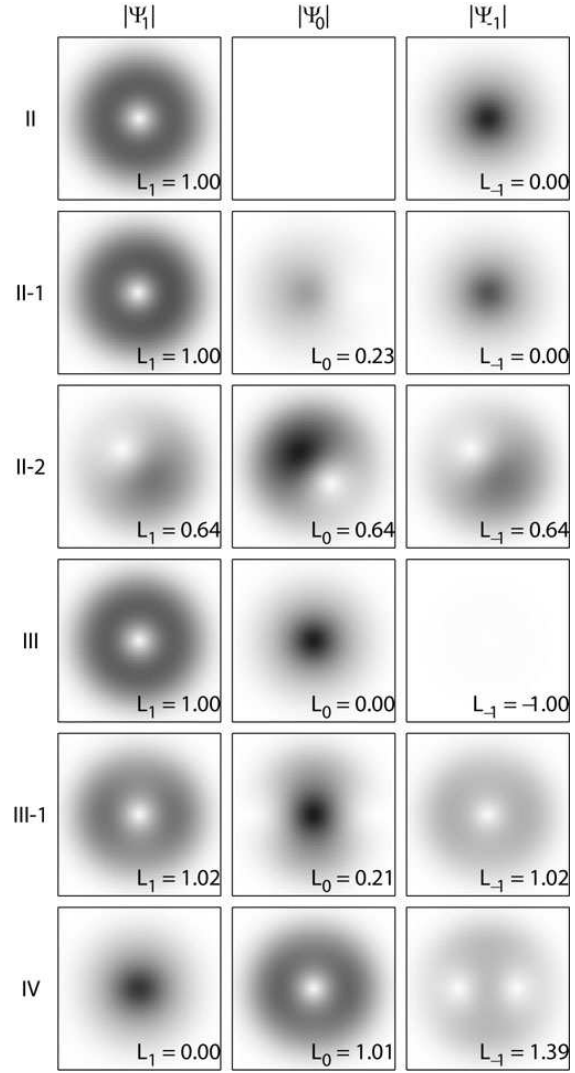}
\end{center}
\caption{\label{fig:2d_plots_eps0.1}
Amplitudes of $|\Psi_i|$ $(i = 0, \pm 1)$ of vortices
in regions
II, II-1, II-2, III, III-1 and IV at $\epsilon=0.1$ and $\Omega = 0.45$.
$L_i$ are the normalized angular moments for each component.
}
\end{figure}

\section{Phase diagrams}

\subsection{Direction of total magnetization}

Total magnetization is conserved in an optically confined condensate
(see refs.~\citen{stenger1998,chang2004}).
For this conservation,
the factor $p$ of the linear Zeeman energy $-pM_z$,
which was originally proportional to the magnetic field,
is treated as a Lagrange multiplier.
Therefore, the factor $p$
loses its direct correspondence with the magnetic field.
Conversely, the quadratic Zeeman factor $q$, which is 
proportional to the square of the magnetic field,
is not restrained by the conservation.

The horizontal axis in the diagrams is the normalized magnetization
$M_z/N$ in the $z$-direction.
The total magnetization $M_x, M_y, M_z$ may be rotated around $z$-axis
by the spin rotation $\exp^{-iF_z \tau}$,
without affecting the Zeeman energies.
Therefore, elements $M_x$ and $M_y$
are denoted by $M_\perp = \sqrt{M_x^2 + M_y^2}$ hereinafter.

The amplitudes of the elements $m_x$ and $m_y$ of the local magnetization,
and therefore the amplitude of $M_\perp$, 
are proportional to the amplitude $|\Psi_0|$ of the zero species.
Because the zero species of the condensate becomes energetically favorable
for larger $q$, 
the total perpendicular magnetization $M_\perp$ can be large for some types of vortices.

Usually, the total magnetization $M_\perp$ perpendicular to 
the $z$-axis is zero.
However, in some regions, vortices with $M_\perp \neq 0$ have the lowest energy.
The regions containing vortices with $M_\perp \neq 0$ are also shown
in Figs.~\ref{fig:eps0.1}(b) and \ref{fig:eps0.01}
as the area shaded with (blue) slant lines.

\subsection{Absence of quadratic Zeeman term}

In the absence of the quadratic Zeeman energy ($q=0$), 
our system is equivalent to that in ref.~\citen{yip1999} at $c_2/c_0 = 0.02$.
In Fig.~\ref{fig:eps0.1}(a), 
regions IV, II and III, which have been reported before~\cite{yip1999},
are observed.
The boundaries of these regions are consistent with the present results.
However, there are also vortex regions II-1 and II-2 exist with $M_\perp \ne 0$ 
and a region III-1 also exists with $M_\perp = 0$.
Figures \ref{fig:eps0.1}(a) and \ref{fig:eps0.1}(b) 
show phase diagrams
along the horizontal axis ($q=0$).
Let us consider these regions from smaller $M_z$ values at $M_\perp = 0$.
Amplitudes of $|\Psi_i|$ of the corresponding vortices are shown in
Fig.~\ref{fig:2d_plots_eps0.1}.

IV, IV-1:
The zero species has a vortex ($L_0 \sim 1$) and 
the other species fill the core region.
Moreover vortex IV-1 has integer
angular momentums $(L_1, L_0, L_{-1}) = (0, 1, 2)$.
For a simple explanation, let us consider the cases 
with a single centered vortex on
one or two of the three components,
and that the other component(s) fillx the vortex core.
For a smaller magnetization $M_z < 0.2$,
the local magnetization $m_z / n = (n_1 - n_{-1})/ n$ tends to be zero.
When the local magnetization $m_z / n$ is zero,
there are two possible configurations:
(1) There is a vortex in the zero component,
and a mixture of up and down species fills the vortex core,
and
(2) the mixture has a vortex (up and down species are co-rotating),
and the zero species fills the core.
Here, under the phase conditions introduced below in eq.~(\ref{eq:phases}),
(1) is preferred.
Hence, type IV is realized.
As $M_z$ is increased from zero, the angular momentum $L_{-1}$ smoothly
increases
to fully satisfy eq.~(\ref{eq:phases}) in the IV-1 area.

III-1:
As mentioned above (2), the up and down species exhibit vorticity
$L_1 \textrm{ and } L_{-1} \sim 1$,
and the zero species fill the core.

II and III:
The total angular momentum around $L = 0.7$ is preferred
due to the angular velocity $\Omega$.
Therefore, the dominant species should have one vortex.
For a larger $M_z$ ($> 0.2$), throughout the II and III regions,
the up component is dominant and has one vorticity $L_1 = 1$.
In region II (III), the down (zero) component fills the core 
to decrease (increase) $M_z$.
The type II and III vortices correspond to
$( 1,\times,0 )$ and $( 1, 0, -1 )$
in ref.~\citen{mizushima2002}, respectively.
In this notation, the (integer) angular moments
of each component are listed in parentheses
for the axisymmetric cases.

Let us consider the vortices with finite $M_\perp$
as shown in Fig.~\ref{fig:eps0.1}(b).
As compared to the corresponding vortex states in Fig.~\ref{fig:eps0.1}(a),
the vortex states in Fig.~\ref{fig:eps0.1}(b) have a lower energy $G$.
The II-1 and II-2 regions are shown in Fig.~\ref{fig:eps0.1}(b),
and these two types of vortices can be smoothly deformed into each other.

From $M_z/N = 0$ to $0.276$: 
For specific value of the parameter $p = 0$,
a configuration of two vortices is possible.
This includes the ``split-(I)" states in ref.~\citen{mizushima2002}.
In our case,
the vortex state has $M_z/N = 0$ and $M_\perp/N = 0.276$.
Because the Zeeman factors are zero, 
the direction of spin may be freely rotated
by the operators $\exp^{-iF_x \alpha}\exp^{-iF_y \beta}$.
By this rotation, this vortex is reduced to
a type II vortex with $M_\perp/N = 0, M_z/N = 0.276$.
By similarly continuing the rotation, these vortices cover a range of
$M_z/N = 0 - 0.276$ while maintaining $\sqrt{M_\perp^2 + M_z^2}/N = 0.276$.
Similar deformations of the density profiles 
among the spin components due to changes in the quantization axis
are reported in refs.~\citen{yip1999,mueller2004}.
This deformation affects the energy
when the Zeeman term factors $p,q$ are finite,
and thus 
no special care is required
in other areas of the phase diagrams.

From $M_z = 0.51$ to $0.66$:
The vortex states in this range
have a small finite perpendicular magnetization $M_\perp/N \sim 10^{-8}$,
and these are similar to the vortices within $0< M_z < 0.276$.
Nevertheless, these cannot be reduced to vortices with $M_\perp = 0$
because of the finite Zeeman term factor $p$.
This region is discussed in \S \ref{sec:mag:perp}.

\subsection{Effects of quadratic Zeeman term}

Let us consider the effects of the quadratic Zeeman energy shown
in eq.~(\ref{eq:e:zee}).
This energy is proportional to the parameter $q$.
An increase in the Zeeman energy
for a finite $q$ is $q(N_1 + N_{-1}) = q(N - N_0)$.
Therefore, there is a maximum increase in the energy of vortex of type II,
whose $N_0 = 0$;
it becomes less favorable.
In the phase diagram, the type II region 
shrinks above $q/(h\nu_r) = 0.05$,
and a large range of $M_z$ is occupied by the other regions.
The other boundaries of the regions are not sensitive to $q$.

The non-axisymmetric region III-1 exists at a finite $q$.
In region III-1,
angular moments $L_1$ and $L_{-1}$ have the same sign and a similar amplitude.
In other words, the up and down species are co-rotating in region III-1,
while they are counter-rotating in region III.
The $c_2$ term of the energy prefers the counter-rotating configuration,
see discussion in \S \ref{sec:mag:perp}.
However, the co-rotating configuration becomes
energetically favorable by decreasing the energy $- \omega L$
because the total angular momentum  
$L \sim (N_1 + N_{-1})/N$ is larger than
$L = (N_1 - N_{-1})/N$ in the counter-rotating case.

Several data points show the III-1 configurations 
even within the region II on Fig.~\ref{fig:eps0.1}(a).
Their energies are lower than the type II vortices.
But we couldn't make sure that 
they (III-1) always exist within certain range of $M_z$
because they become unstable by small changes in $p$.

\begin{figure}
\begin{center}
\begin{tabular}{cc}
\includegraphics[width=4.2cm,clip]{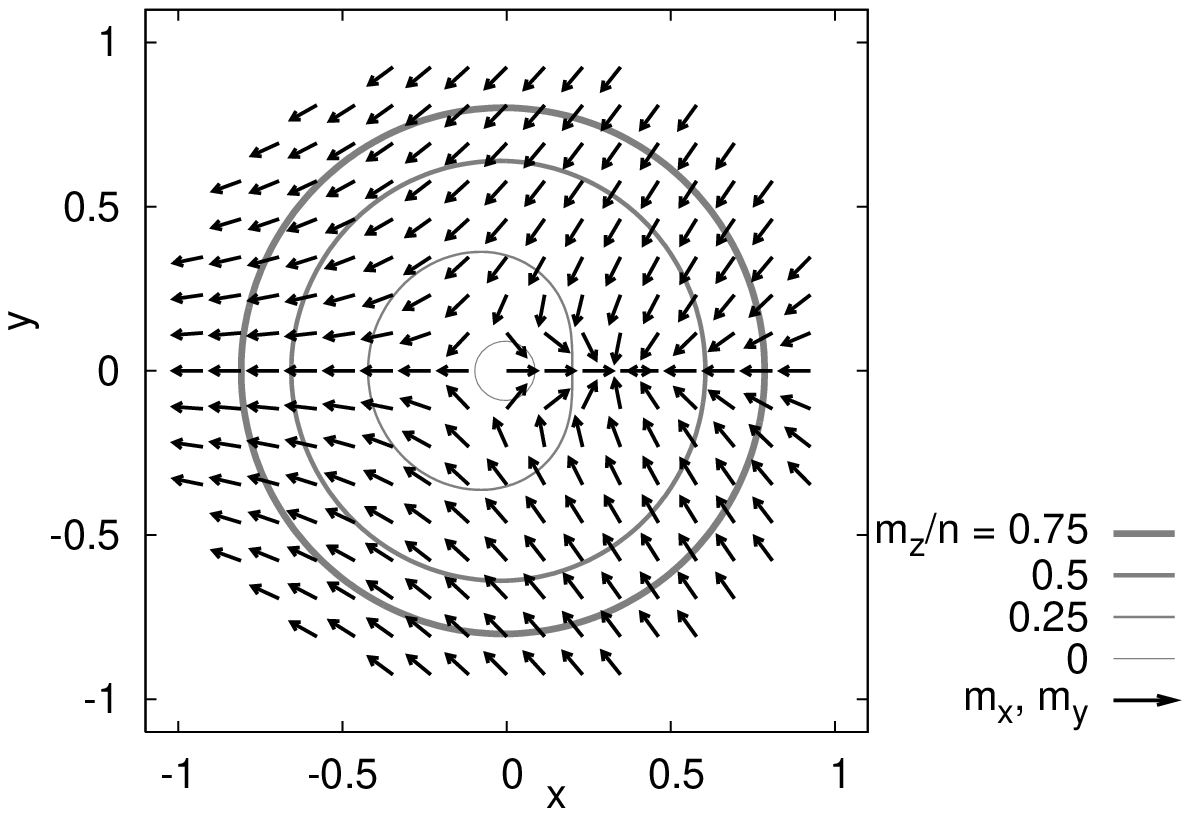} &
\includegraphics[width=3.1cm,clip]{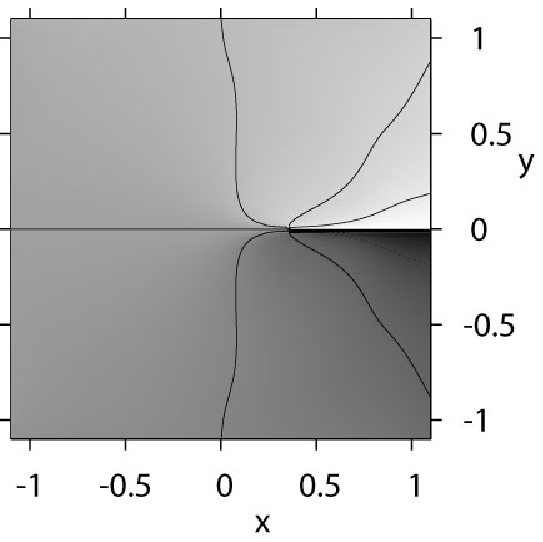}
\\
(a) & (b)
\\
\includegraphics[width=4.2cm,clip]{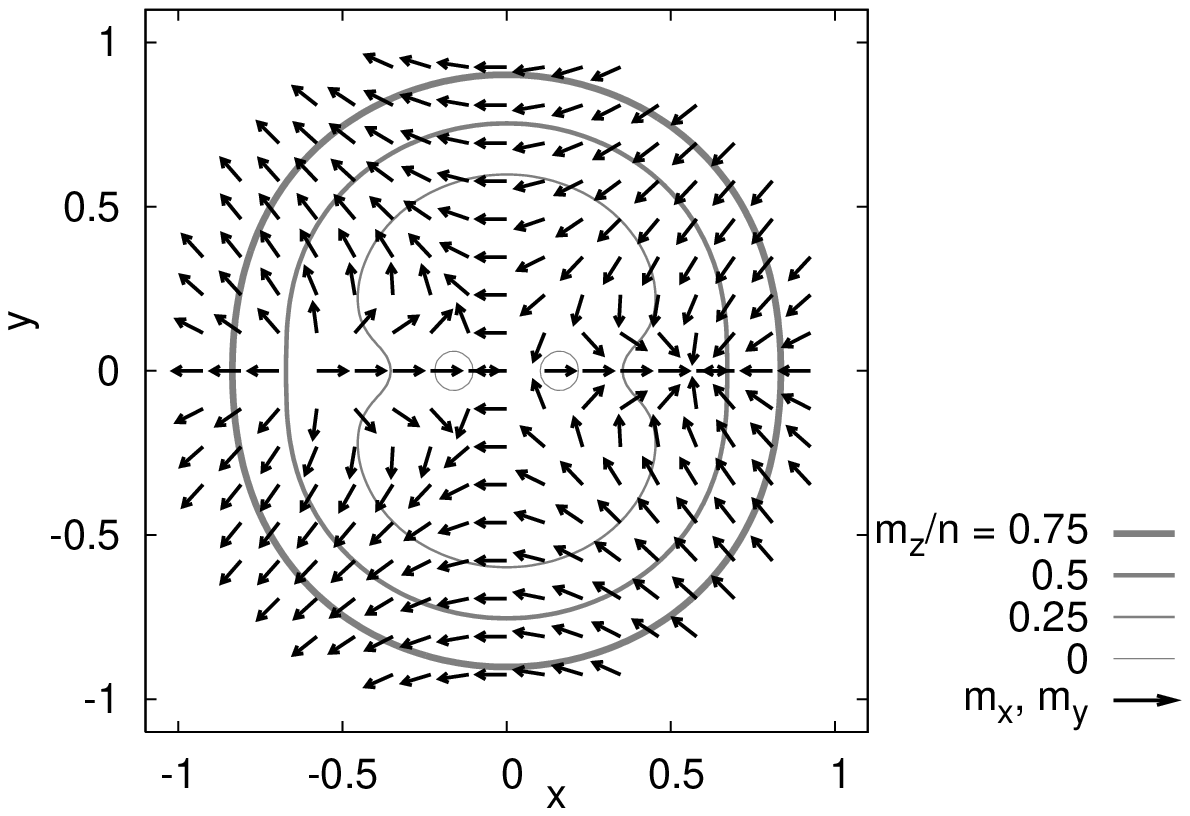} &
\includegraphics[width=3.1cm,clip]{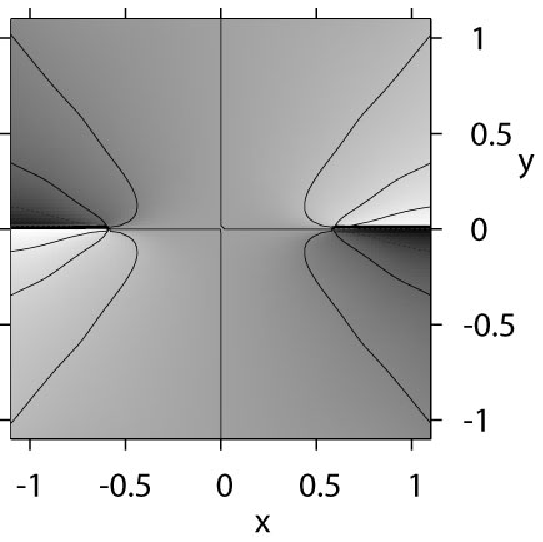}
\\
(c) & (d)
\\
\end{tabular}
\end{center}
\caption{\label{fig:arrow}
(a)(b) A vortex state within the region II-1
at $\varepsilon = 0.01$, $q = 0.06 h\nu_r$, $M_z/N = 0.437$ and $M_\perp/N = 0.080$.
(c)(d) A vortex state within the region III-2
at $\varepsilon = 0.01$, $q = 0.05 h\nu_r$, $M_z/N = 0.302$ and $M_\perp/N = 0.085$.
(a)(c) Spin textures.
Arrows show the directions of local spin components $m_x$ and $m_y$.
Component $m_z/n$ is shown by contours.
(b)(d) Phases of $\Psi_0$. Solid lines indicate multiples of $\pi/4$.
}
\end{figure}

\subsection{Phase in the region II-1}\label{sec:mag:perp}

In the region II-1, 
the phases and angular momentums
of the up and down species are similar to those in the region II.
However, the wavefunction $\Psi_0$ is nonzero there.
This minor change in the wavefunction significantly 
changes the spin texture
because the components of local magnetization $m_x$ and $m_y$ are
proportional to the amplitude $|\Psi_0|$.
The spin texture in the $xy$ plane appears
as shown in Fig.~\ref{fig:arrow}(a).
The phase of $\Phi_0$ is important for understanding of this spin texture.

The phase of $\Psi_0$ is restricted by
the minimization of the local spin interaction energy
of $E_\mathrm{int}$ in eq.~(\ref{eq:e:int}).
The $c_2$-proportional part of the local energy is written as
\begin{eqnarray}
    \mathcal{E}_{c2} &=& 
    \frac{c_2}{2} n + 
    \frac{c_2}{2} \left(
         n^2 - |2 \Psi_1\Psi_{-1} - \Psi_0^ 2|
    \right)
\label{eq:ec2}
\\
    &=& 
    C +
    2 c_2 \sqrt{n_1 n_{-1}} n_0 \cos(2\alpha_0 - \alpha_1 - \alpha_{-1})
\end{eqnarray}
where $\alpha_j$ are the phases of $\Psi_j$,
and $C$ is a function of densities $n_j$.
In our antiferromagnetic system ($c_2 > 0$),
a condition for minimizing $\mathcal{E}_{c2}$ for given $n_j$ is 
%
%
\begin{equation}
    2\alpha_0 - \alpha_1 - \alpha_{-1} = \pi + 2m\pi,
\label{eq:phases}
\end{equation}
%
%
where $m$ is an integer.

In regions II and II-1, $\alpha_1 \simeq \theta, \alpha_{-1} = 0$.
Therefore, $\alpha_0 = \frac{\theta}{2}$
and the zero species must have a half winding number.
In region II-1,
instead of a half winding number,
the zero species has an off-centered vortex core
and a non-integer angular momentum $0 < L_0 < 1$, see Fig.~\ref{fig:arrow}(b).


The components of the magnetization $m_x$ and $m_y$ are
proportional to the amplitude $|\Psi_0|$.
In II-1, because the amplitude $|\Psi_0|$ becomes finite,
the spin texture in the $xy$ plane appears
as shown in Fig.~\ref{fig:arrow}(a).

In the absence of Zeeman terms $p=q=0$, this structure
may be reduced to vortices in region IV by changing the quantization axis.
At a finite $p$ or $q$, these vortices cannot be reduced similarly
and
are energetically more favorable than type II vortices.
Therefore, the II-1 vortex is
likely to be observed in experiments conducted with finite magnetic fields.

\begin{figure}
\begin{center}
\includegraphics[width=7.5cm,clip]{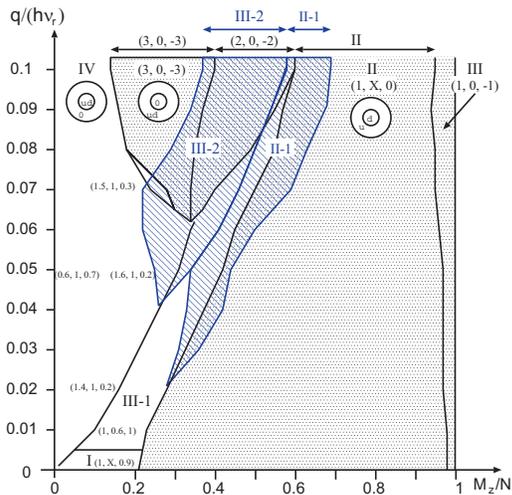}
\end{center}
\caption{\label{fig:eps0.01}(Color online)
Phase diagram of magnetization $M/N$ vs quadratic Zeeman term $q/(h\nu_r)$
at $\epsilon=0.01$ and $\Omega = 0.07$.
Notations are same as in Figs.~\ref{fig:eps0.1}(a) and \ref{fig:eps0.1}(b).
But both $M_\perp = 0$ regions and 
$M_\perp \ne 0$ regions (shaded with slash lines) are shown here.
A region at $q=0.07, M_z/N \sim 0.25$ is not labelled.
There angular momentums are around $(2.7,0.3,0.3)$.
}
\end{figure}
\begin{figure}
\begin{center}
\includegraphics[width=7.4cm,clip]{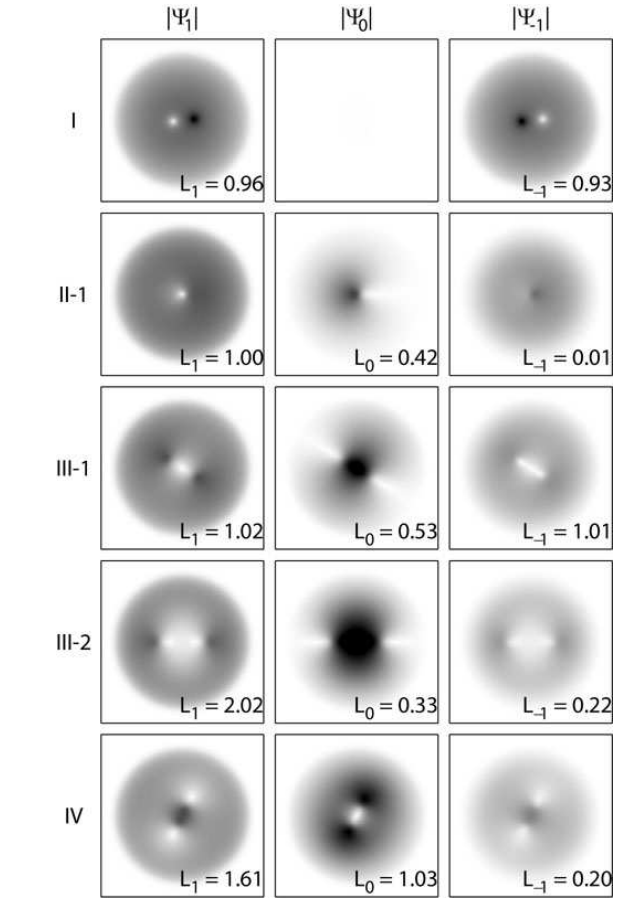}
\end{center}
\caption{\label{fig:2d_plots_eps0.01}
Amplitudes of $|\Psi_i|$ $(i = 0, \pm 1)$ of vortices in regions
I, II, III, III-1, IV and IV-1 at $\epsilon=0.01$ and $\Omega = 0.07$.
The normalized angular momentum $L_i$ is shown in each panel.
}
\end{figure}

\subsection{Disk-shaped condensate} 

Figure \ref{fig:eps0.01} is the phase diagram as $\varepsilon = 0.01$.
Amplitudes of $|\Psi_i|$ of the corresponding vortices are shown in
Fig.~\ref{fig:2d_plots_eps0.01}.
This value corresponds to
the realistic particle number for tight $z$-confinement traps.
We employ the angular velocity $\Omega = 0.07 \times 2 \pi$
to cover the vortex states observed at $\varepsilon = 0.1$.
At a lower (higher) rotation, $\Omega = 0.05\times 2 \pi$ ($0.1\times 2 \pi$), 
vortexfree (many vortex) states replace these vortex states.

At $q=0$, regions I, II and III are observed.
Region I is reported in ref.~\citen{yip1999}.
The composite vortex in this region consists of two vortices,
one is in the up component and the other is in the down component.
The zero component does not have an atom.
The composite vortex in region I at $p = q = 0$ is
equivalent to that in region IV
through the $\pi/2$ rotation of the quantization axis.
The other regions III and IV are similar to the abovementioned
$\varepsilon=0.1$ case.

At finite $q$, because the absence of the zero component is energetically
unfavorable, region I is replaced by region III-1.
The vortices in region III-1 may be understood as
those in region I with finite atoms of zero species.

Regions II, III, III-1 and IV are similar to the $\varepsilon=0.1$ case.
However, the region II elongates toward larger $q$
because the scale of $\varepsilon$ changes with $q$.
Consider two energy densities at the trap center
$\mathcal{E}_{c2} \sim c_2 n^2 \propto  c_2\varepsilon^{-2}$
and 
$\mathcal{E}_q \sim q n \propto q\varepsilon^{-1}$,
where $n$ is the atom density.
These energies determine the relative population of the zero species,
and they should be comparable at the boundary of region II.
When we consider the boundary between systems with different $\varepsilon$
values,
systems with equal $q\varepsilon$ rather than $q$
are in correspondence.

In regions III, $(2,0,-2)$ and $(3,0,-3)$,
the up and down species are counter-rotating with respect to each other
(angular moments $L_1$ and $L_{-1}$ have opposite sign but the same amplitude).
These satisfy the phase relation given by eq.~(\ref{eq:phases}).
In region III-1,
this condition is not met.
Region III-1, in which the up and down species are co-rotating,
lies in between regions IV and II.

Vortices with finite perpendicular magnetization $M_\perp$ 
reduce the energy
at larger $p$.
They are shown as areas shaded with oblique lines,
\textit{e.g.}, II-1 and III-2 regions.
The II-1 region at $M_z/N \sim 0.5$ that also appeared when $\epsilon=0.1$.
Here, $M_\perp/N \sim 0.03 - 0.09$.
Another region is III-2, in which $M_\perp/N \sim 0.07$.
A pair of spin structures similar to those 
of the type II-1 vortices exist at $\varepsilon = 0.1$.

When the initial perpendicular magnetization is zero ($M_\perp/N = 0$)
and the total magnetization is explicitly conserved,
only the vortex configurations with $M_\perp/N = 0$ are selected.
If there is an error in $M_\perp/N$, the II-1 or III-2 type of vortices
having lower energy will be realized.
Therefore, determining whether the areas shaded by oblique lines
should be considered in this phase diagram
depends on the experimental details.

The vortices in region III-2 
have a pair of the line-like phase singularities on the 
zero species
as shown in Fig.~\ref{fig:arrow}(d).
Figure \ref{fig:arrow}(c) presents resulting spin texture.
Between the singularities, the center of the trap is filled by zero species.
These vortices are regarded as a pair of type II vortices 
that has gained finite zero species $\Psi_0$.

\section{Conclusion}

We investigated the effect of the quadratic Zeeman energy
on various vortex states of an antiferromagnetic spinor BEC.
We presented phase diagrams 
showing the quadratic Zeeman term factor $q$ vs 
magnetization $M_z/N$.
An important observation is that 
the perpendicular magnetization $M_\perp/N$
is also necessary to correctly plot the phase diagram,
not only at finite $q$ but also at $q=0$.
The permissible range of $M_\perp$ will depend on the experimental conditions.

In a specific case ($p=q=0$) in which the Zeeman terms are ignored,
the rotation of all the spins 
deforms a vortex in region II into other configurations (regions II-1 and II-2)
with a finite perpendicular magnetization $M_\perp$.
This covers the range $0 \le M_z/N \le 0.276$
while maintaining $\sqrt{M_z^2 + M_\perp^2}/N = 0.276$.

For a finite $q$, due to the quadratic Zeeman energy,
vortices with $M_\perp \ne 0$ (in regions II-1, II-2 and III-2)
widely replace the other vortices and change the phase diagram.
In the general case ($p \ne 0$ or $q \ne 0$),
configurations with finite $M_\perp$
cannot be reduced into configurations with $M_\perp = 0$.
Axisymmetric vortices $(3,0,-3)$ and $(2,0,-2)$
whose up and down components are counter-rotating with respect to each other
are also found at larger $q$.

In the absence of the quadratic Zeeman term ($q=0$),
the result is consistent with previous studies by one of us~\cite{yip1999}.
However, vortices II-1 and III-1 are also found.
These vortices may be understood as a deformation of the vortices
in regions II and I with finite zero species.

The vortices in region II are commonly known for the antiferromagnetic
spinor BEC~\cite{martikainen2002,mizushima2002,mueller2004}. 
For a finite $q$, these vortices gains zero species and
are deformed into II-1 vortices.
Similarly, III-2 vortices are regarded as derived from a pair of type II vortices.
A similar deformation of vortices by having finite zero component
is expected in the spinor BEC in a magnetic field,
especially in the lattices of the type II vortices~\cite{mueller2004}.


\end{document}